\documentclass[epj,twocolumn]{webofc}
\usepackage[varg]{txfonts}   
\usepackage{graphicx}

\wocname{epj}
\woctitle{Seismology of the Sun and the Distant Stars 2016}
\begin{document}

\title{Effect of \emph{Kepler} calibration on global seismic and background parameters}
%
\subtitle{Preliminary study on a sample of seismic solar analogs}

\author{\firstname{David} \lastname{Salabert}\inst{1}\fnsep\thanks{\email{david.salabert@cea.fr}} \and
        \firstname{Rafael A.} \lastname{Garc\'ia}\inst{1}\and
        \firstname{Savita} \lastname{Mathur}\inst{2}\and
        \firstname{J\'er\^ome} \lastname{Ballot}\inst{3,4}
}

\institute{Laboratoire AIM, CEA/DRF-CNRS, Universit\'e Paris 7 Diderot, IRFU/SAp, Centre de Saclay, 91191, Gif-sur-Yvette, France 
\and
          Center for Extrasolar Planetary Systems, Space Science Institute, 4750 Walnut street Suite\#205, Boulder, CO 80301, USA
\and
          CNRS, Institut de Recherche en Astrophysique et Plan\'etologie, 14 avenue Edouard Belin, 31400, Toulouse, France
\and
	Universit\'e de Toulouse, UPS-OMP, IRAP, 31400, Toulouse, France
}

\abstract{
Calibration issues associated to scrambled collateral smear affecting the {\it Kepler} short-cadence data were discovered in the Data Release 24 and were found to be present in all the previous data releases since launch. In consequence, a new Data Release 25 was reprocessed to correct for these problems.  We perform here a preliminary study to evaluate the impact on the extracted global seismic and background parameters between data releases. We analyze the sample of seismic solar analogs observed by {\it Kepler} in short cadence between Q5 and Q17. We start with this set of stars as it constitutes the best sample to put the Sun into context along its evolution, and any significant differences on the seismic and background parameters need to be investigated before any further studies of this sample can take place. We use the A2Z pipeline to derive both global seismic parameters and background parameters from the Data Release 25 and previous data releases and report on the measured differences. 
}
\maketitle
%
\section{Introduction}
\label{intro}
The study of the characteristics of solar analogs \cite{cayrel96} of different ages is a very promising way to understand the evolution of the Sun. It offers as well the possibility to infer new constraints on the variability of the solar magnetic activity and its associated dynamo processes. However, the identification of solar-analog stars depends on the accuracy of the estimated fundamental stellar parameters. The unprecedented quality of the continuous four-year photometric observations collected by the {\it Kepler} satellite \cite{borucki10} allowed the measurements of acoustic oscillations in hundreds of solar-like stars \cite{chaplin14}. The addition of asteroseismic data combined with high-resolution spectroscopic observations was proven to provide the most accurate fundamental properties that can be derived from stellar modeling today, either from global oscillation properties \cite{chaplin14} or from individual frequencies \cite{mathur12, metcalfe14,silva15,creevey16} compared to empirical scaling-law relations \cite{kjel95}. In consequence, the selection of stars analog to the Sun will benefit from the inclusion of the detection of solar-like oscillations as an additional selection criterium. This is what we called a seismic solar-analog star \cite{salabert16a}, thus extending the classical definition \cite{cayrel96}. Among the {\it Kepler} sample of oscillating solar-like stars, we identified 18 seismic solar analogs \cite{salabert16a}.

However, calibration issues\footnote{See detailed explanations at \url{http://archive.stsci.edu/kepler/}.} present since the launch of {\it Kepler} were discovered in the Q2--Q16 Data Release 24 (DR24) short-cadence (SC) data, affecting approximately half of all {\it Kepler} SC targets. We note that the long-cadence (LC) data are not affected. The origin of the problems was traced to an accounting error which scrambled much of the SC collateral smear data used to correct for the effects of {\it Kepler}'s shutterless readout. 
As a result, the available SC calibrated pixels, and the light curves based on these pixels, are noisier. The final reprocessing of the {\it Kepler} Q0--Q17 SC light curves corrected from these calibration problems was thus performed and made available with DR25 in summer 2016. 

It is thus important to comprehend the impact of the calibration issues affecting the previous releases of {\it Kepler} data, and the differences and the possible bias with DR25 on the derived stellar parameters. Here, we present a preliminary study on the differences between DR25 and previous releases on the global seismic and background parameters applied for the sample of 18 solar-analog stars [9]. Although the studied sample is rather small to perform a thorough statistical analysis, we provide initial inputs on the expected effects. Furthermore, this sample constitutes the set of stars the most comparable to the Sun observed by {\it Kepler} and any possible inferences on the estimated stellar parameters due to issues in the data calibration should be known for each target before any further studies of this sample are performed.

\begin{figure*}
\centering
\includegraphics[width=1\textwidth,clip]{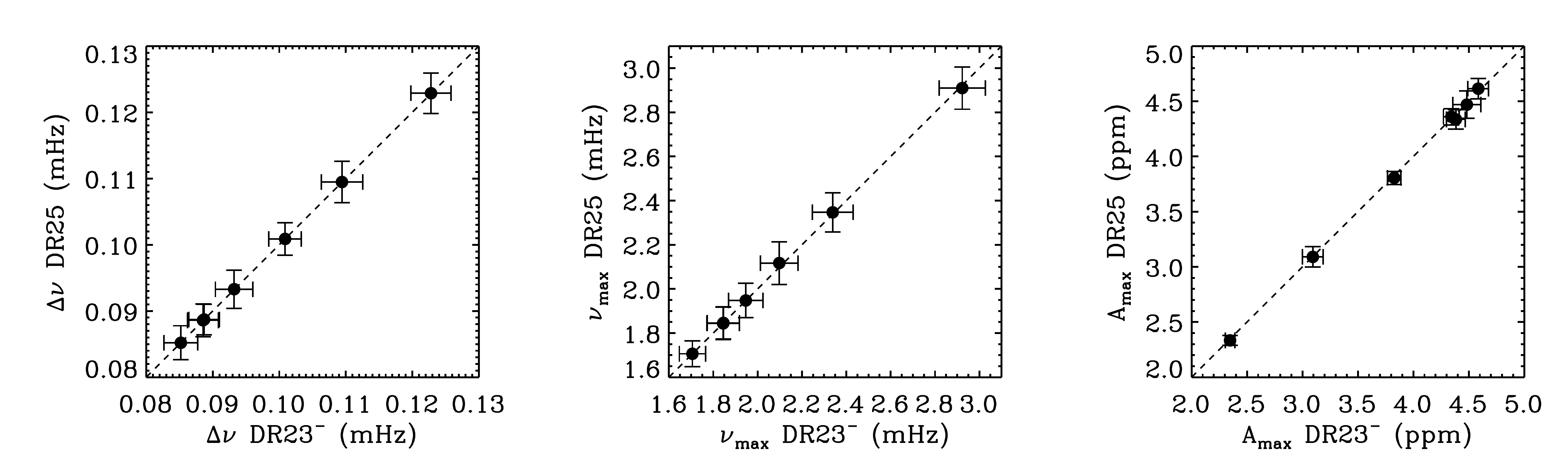}
\caption{Comparison of the derived global seismic parameters between {\it Kepler} short-cadence observations from DR23$^-$ and DR25. {\it From left to right}: the mean large separation $\Delta\nu$, the frequency of maximum power $\nu_\mathrm{max}$, and the maximum amplitude $A_\mathrm{max}$ at $\nu_\mathrm{max}$. In the three panels, the dashed lines correspond to the 1:1 relation. The associated uncertainties are also represented.}
\label{fig-1}       
\end{figure*}

\section{Global seismic and background parameters}
\label{sec-2}
Out of the 18 seismic solar analogs observed with {\it Kepler} \cite{salabert16a}, nine of them were observed in short cadence from Q5 to Q17: KIC\,3656476, KIC\,4914923, KIC\,5774694, KIC\,6116048, KIC\,7296438, KIC\,7680114, KIC\,9098294, KIC\,10644253, and KIC\,11127479. We checked that all these targets were affected by the  smear calibration issues from the list of affected stars available at the MAST archive\footnote{See list of affected targets at \url{https://archive.stsci.edu/missions/kepler/catalogs/kepler_scrambled_short_cadence_collateral_target_list.csv.}}. We note that as the two stars KIC\,5774694 and KIC\,11127479 were observed only during one and three quarters respectively, they were disregarded for the following analysis. A total of seven solar analogs were then analyzed. DR25 and previous data releases (DR21 or DR23 depending on the star, denoted in the rest of the text DR23$^-$) were corrected for the usual issues found in the {\it Kepler} light curves following the procedures explained in \cite{garcia11}. The gaps present in the data and affecting the stellar background were filled using inpainting methods \cite{garcia14,pires15}.

The global seismic parameters of our sample of stars were determined using the A2Z pipeline \cite{mathur10}. The mean large frequency, $\Delta \nu$, which is the distance between two modes of same angular degree and consecutive radial order, was measured by estimating the highest peak in the power spectrum of the power spectrum. The stellar background parameters were extracted by fitting the power spectrum with a function based on the results from \cite{kallinger14}. In summary, five different components were fitted to the power spectrum: a white instrumental noise, two Harvey-like functions to mimic two granulation contributions \cite{harvey93}, a power-law for the magnetic activity, and a Gaussian function for the p-mode power-excess envelop. The slopes of the Harvey-like functions were fixed to four as it was demonstrated to provide a better fit \cite{kallinger14}. The parameters of the Gaussian function provide the center of the p-mode bell, which corresponds to the frequency of maximum power, $\nu_{\rm max}$. The associated maximum amplitude, $A_\mathrm{max}$, is also returned.

\subsection{Global seismic parameters}
Figure~\ref{fig-1} shows the global seismic parameters ($\Delta\nu$, $\nu_\mathrm{max}$, and $A_\mathrm{max}$) extracted from both DR23$^-$ and DR25 for the seven analyzed solar analogs. The corresponding uncertainties are also represented. No systematic bias can be seen between DR23$^-$ and DR25, and the differences in the global seismic parameters are minimal (less than 1\%) and well within the uncertainties.

\begin{figure}
\centering
\includegraphics[width=0.45\textwidth,clip]{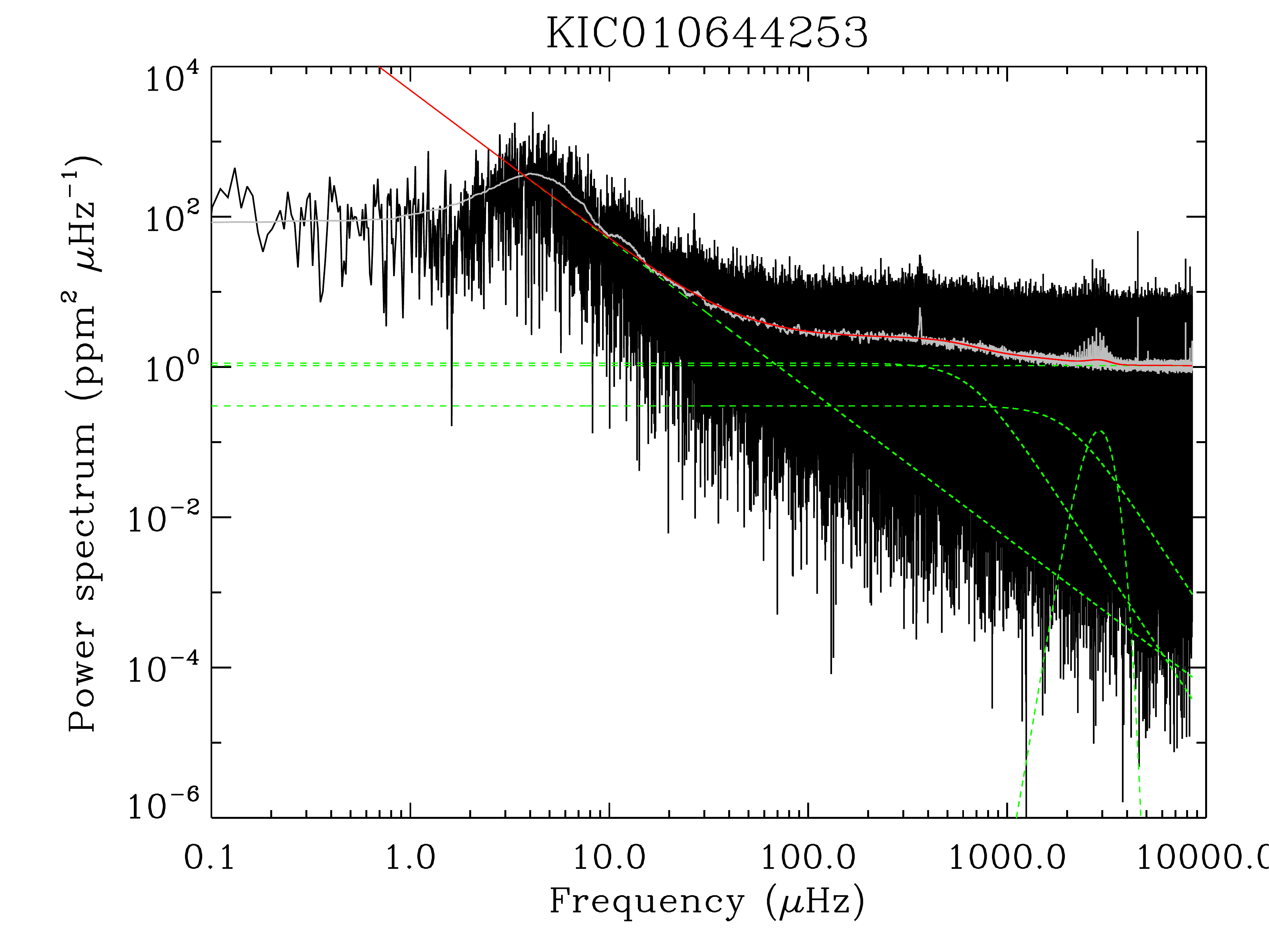}
\caption{Power spectrum of the solar analog KIC\,10644253 observed by {\it Kepler} DR23 (in black without smoothing, in grey with smoothing). The green dashed lines correspond to the different fitted components of the background model. The red solid line corresponds the fitted background model.}
\label{fig-2}     
\end{figure}

\begin{figure*}
\centering
\includegraphics[width=1\textwidth,clip]{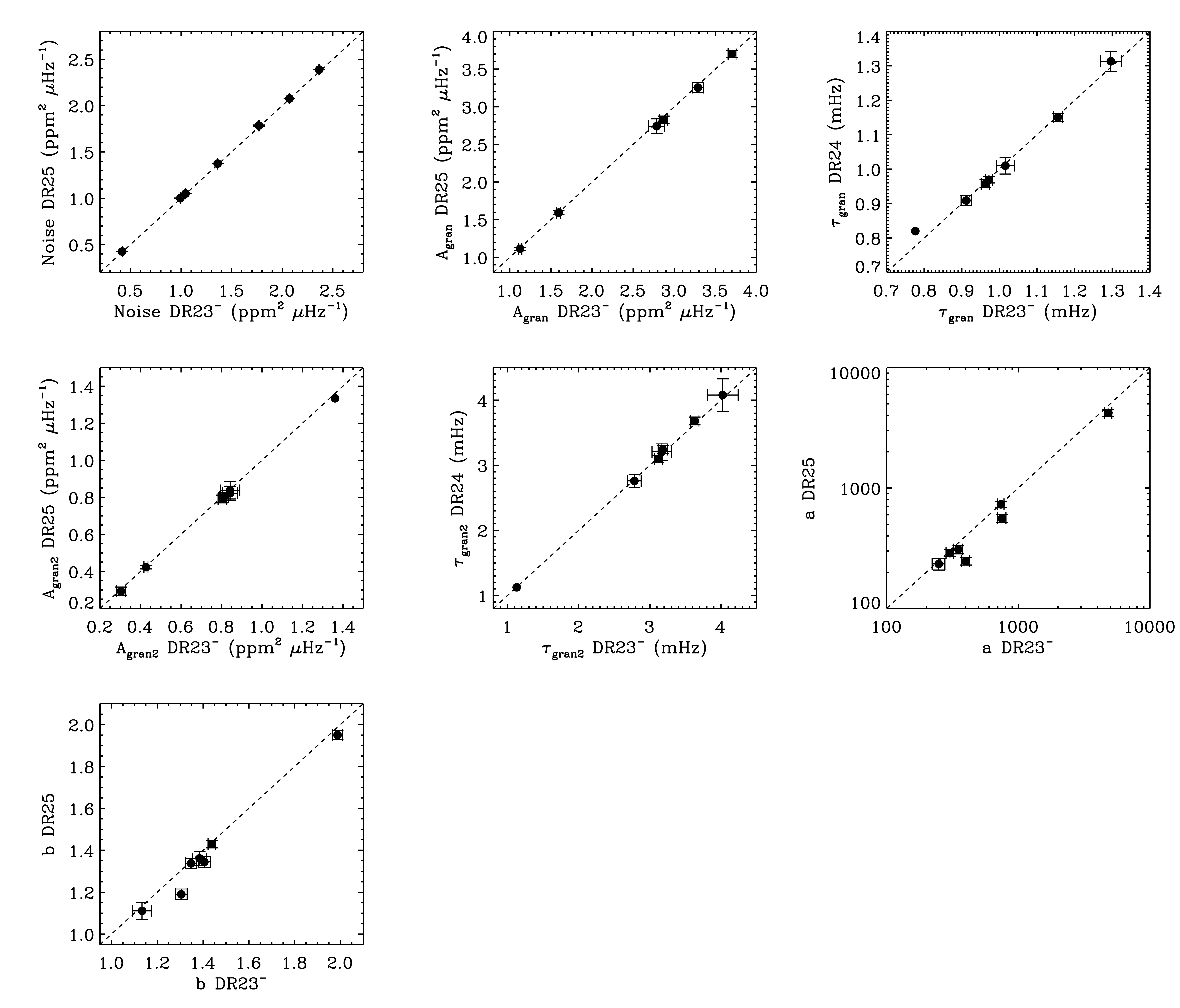}
\caption{Comparison of the derived stellar background parameters between {\it Kepler} short-cadence observations from DR23$^-$ and DR25. In the three panels, the dashed lines correspond to the 1:1 relation. The associated uncertainties are also represented.}
\label{fig-3}       
\end{figure*}

\subsection{Background parameters}
An example of the background model used here is illustrated on Fig.~\ref{fig-2} in the case of the 1-Gyr-old solar analog KIC\,10644253 from DR23. This star is a particularly interesting target as it is the youngest G-type star within the solar-like oscillators observed by {\it Kepler} [5,~9]. Furthermore, it exhibits temporal variations of the oscillation frequencies associated to its magnetic variability \cite{salabert16b}. Such temporal variability of the acoustic oscillation parameters are well studied in the case of the Sun over to date two 11-year solar cycles \cite[see e.g.,][and references therein]{salabert15}, but has been observed so far in only three stars: the F-type stars HD\,49933 \cite{garcia10,salabert11} and KIC\,3733735 \cite{regulo16}, and the solar-analog G-type star KIC\,10644253 [13].

The different components of the background model (i.e., the two Harvey's laws, the p-mode gaussian envelop, the power law, and the white instrumental noise) are represented individually as well on Fig.~\ref{fig-2}. The background fitting was performed for the sample of solar-analog stars and each of the background components were compared between DR23$^-$ and DR25 as shown on Fig.~\ref{fig-3}. The corresponding uncertainties are also represented. The granulation components, the amplitudes, $A_\mathrm{gran}$ and $A_\mathrm{gran2}$, and the characteristic frequencies, $\tau_\mathrm{gran}$ and $\tau_\mathrm{gran2}$, and the instrumental noise do not show significant differences between DR23$^-$ and DR25 (respectively 3\% and 1\%). However, the two parameters, $a$ and $b$, of the power law $a\nu^{-b}$ describing the stellar activity and low-frequency trends show significant differences for some of the stars between DR23$^-$ and DR25 up to 60\% and 10\% respectively. Furthermore, all the parameters $a$ and $b$ are smaller when derived with DR25 compared to DR23$^-$. It appears then from this small sample that the principal and significant impact of the smear calibration issues concerns the parameters of the power law in the background model.

\section{Conclusions}
\label{sec-con}
Calibration problems associated to collateral smear were detected  in the {\it Kepler} DR24 and reprocessed corrected light curves were provided in DR25. These issues were actually found to affect all data releases since launch. We performed here a preliminary study to evaluate the impact on the global seismic parameters and on the background parameters between the two data releases. We analyzed the sample of seven seismic solar analogs observed by {\it Kepler} in short cadence continuously between Q5 and Q17. We started with this set of stars as it constitutes the best sample to put the Sun into context along its evolution, and any significant differences on the seismic and background parameters need to be investigated before any further studies of this sample take place. 

We used the A2Z pipeline to derive both global seismic parameters and background parameters with DR23$^-$ and DR25. We found that the global seismic parameters, i.e. the mean large frequency separation $\Delta\nu$, the frequency of maximum power $\nu_\mathrm{max}$, and the maximum amplitude $A_\mathrm{max}$, do not show significant differences (less than 1\%) within the uncertainties. The components of the background model used to describe the granulation (the amplitudes $A_\mathrm{gran}$ and $A_\mathrm{gran2}$, and the characteristic frequencies $\tau_\mathrm{gran}$ and $\tau_\mathrm{gran2}$) with two Harvey's empirical laws are consistent between DR23$^-$ and DR25 within less than 3\% as well as the white instrumental noise (1\%). However, significant differences are observed for some of the stars in the power-law parameters, $a$ and $b$, which are used to describe the response of stellar activity and low-frequency trends in the power spectrum. These parameters are smaller in DR25 by up to 60\% and 10\% for $a$ and $b$ respectively. 

Although the number of analyzed stars is small, it appears that a particular attention is, at least, needed when studying the parameters extracted from the low-frequency power-law fitting. The analysis of a larger sample of stars is still necessary to get a better idea of the differences between DR25 and previous data releases, but this study presented here provides some initial elements  on what can be expected. Furthermore, some additional individual oscillation modes are possibly observed at low frequency, which would provide additional constraints in modeling of stellar parameters \cite{mathur12,metcalfe14,silva15,creevey16}. Peak-fitting analysis of the individual modes with DR25 is thus necessary to confirm this possibility.
\\\\

\section*{Acknowledgments}
{The authors wish to thank the entire {\it Kepler} team, without whom these results would not be possible. Funding for this Discovery mission is provided by NASA's Science Mission Directorate. DS and RAG acknowledge the financial support from the CNES GOLF and PLATO grants. SM acknowledges support from the NASA grant NNX12AE17G and NNX15AF13G.}\\

%
%

\end{document}